\documentclass[%
 reprint,
 amsmath,amssymb,
 aps,
]{revtex4-1}

\usepackage{graphicx}% Include figure files
\usepackage{dcolumn}% Align table columns on decimal point
\usepackage{bm}% bold math
\begin{document}

\preprint{NSTec 25946--2616}

\title{Coupling Multiple SU(2) Symmetry Groups Into A SU(3) Symmetry Group}

\author{Mark L. A. Raphaelian}
\affiliation{National Security Technologies, LLC., Livermore CA 94550 }%
 \altaffiliation[Also at ]{https://www.linkedin.com/in/markraphaelian}
 \email{email: raphaeml@nv.doe.gov}

\date{\today}

\begin{abstract}
A specific algebraic coupling model involving multiple quantization axes is presented in which previously indistinguishable SU(2) symmetry groups become distinguishable when coupled into a SU(3) group structure. The model reveals new intrinsic angular momentum (or isospin) eigenvectors whose structural symmetries are detailed, some of which are not available for groups having only one quantization axis available for configurations. Additionally, an intrinsic cyclic ordering of the quantization axes and internal geometric phase relations between the basis states naturally fall out from the algebraic properties of the group. The nature of the coupling into a SU(3) symmetry group also allows for the definition of both positive and negative basis states along each quantization axis relative to the center of the group. These basis states configurations mimic those observed in color-charge and electroweak theories.
%\begin{description}
%\item[PACS numbers]{02.20.-a, 03.65.Fd, 03.65.Vf, 11.10.Nx, 11.10.St, 11.30.Ly, 11.30.Na, 12.90+b, 14.80.-j}
%\end{description}
\end{abstract}

% PACS, the Physics and Astronomy Classification Scheme.
%\keywords{Suggested keywords}
%Use showkeys class option if keyword display desired
\maketitle

\section{Introduction}

The three SU(2) groups contained within the SU(3) group have been known for many years now \cite{Meshkov1,Han1}. The main physics related conceptional aspects associated with these groups within SU(3) are: nuclear models \cite{Elliot1}, flavor symmetries by way of I-spin, U-spin, and V-spin doublets \cite{Gell-Mann1,Fritzsch1}, and ``color-charge'' associated with quarks \cite{Fritzsch2,Bardeen1}. Even though the details of nuclear models and flavor spin doublets are valid in every aspect, the ``color-charge'' aspects associated with this group are said to be an exact symmetry.

While the three-dimensional SU(3) group may be familiar to many of us, the underlying physics associated with the group may not be. The physics is definitely is not as familiar as what we know and associate with the two-dimensional SU(2) group. The extra column and row that the SU(3) matrices contain make the intrinsic understanding of the underlying physics more obscure and less easily to comprehend. This paper attempts to alleviate this issue by simplifying the physics contained within the SU(3) group into that associated with three coupled SU(2) groups. In doing so we can access and apply the foundations associated with the physics of the SU(2) group to understanding the physics and structures associated with and contained within the SU(3) group.

This first part of this paper explores the algebraic details of the SU(3) group and the associated coupling of multiple SU(2) symmetry axes with one another into a multi-quantization axis group structure. Specifically, the SU(3) group vector space is defined by and is solely comprised by three coupled SU(2) groups, each having an associated matrix group that includes an independent quantization axis. This algebraic foundation allows for the definition of three intrinsic angular momentum (or iso-spin) groups with basis states along each quantization axes. These basis states form single-axis and multi-axis eigenvector configurations that have very specific exchange transformation symmetries. Configurations involving one, two, three, and more coupled basis states are presented.

This second part of this paper explores the geometrical structure of the SU(2) groups within SU(3) group that allows for the realization of definable phase relations between the quantization axes and basis states. These phases are internal to the coupled group and are referenced to the positive directionality or negative directionality associated with each quantization axis. Raising and lowering operations allow for the coupling of basis state configurations defined on the same or different axes in addition to the coupling between positive directionality defined basis state configurations and negative directionality defined basis state configurations.

\section{The Algebraic Relationships Of SU(2) Subgroups Within SU(3)}

The SU(3) algebraic group is defined by the Gell-Mann matrices $\lambda_{i}$ where the subscript runs from 1 to 8. These three dimensional matrices contain three SU(2) groups coupled together by very specific algebraic relationships to form the SU(3) group. This section defines these relationships.

Six of the Gell-Mann matrices contain only off diagonal elements. They are $\lambda_{1}$, $\lambda_{2}$, $\lambda_{4}$, $\lambda_{5}$, $\lambda_{6}$, and $\lambda_{7}$. Two of the matrices, $\lambda_{3}$ and $\lambda_{8}$, contain only diagonal elements and are called the generators of the group. These matrices represent a set of complex vectors with the latter two matrices also representing two orthogonal quantization axes that define a plane. On this two-dimensional plane the SU(3) eigenvectors produce two SU(3) eigenvalues. One lying on the quantization axis produced by the $\lambda_{3}$ generator and the other lying on the quantization axis produced by the $\lambda_{8}$ generator.

For the SU(3) group's quantum numbers to be solely comprised from the coupling of three SU(2) groups, then the three groups SU(2) quantization axes must also lie on the plane defined by the SU(3) group quantization axes. To extract the three SU(2) symmetry axes contained within the SU(3) group, we construct three ``0-component'' matrices. Each of the three ``0-component'' matrices represent one the SU(2) quantization axes, (12), (45), or (67), found in the SU(3) group. The three coplanar ``0-component'' matrices are defined as:
%\begin{eqnarray}\nonumber\begin{array}{cl}
\begin{eqnarray}\begin{array}{cl}
\label{zero-component}
\lambda^{12}_{_{0}}\,\,=\,\,\lambda_{3}, & \qquad \lambda^{45}_{_{0}}\,\,=\,\,-\frac{1}{2}\left(+\,\lambda_{3} \,+\,\sqrt{3} \lambda_{8}\right), \\ \\ \mathrm{and} & \qquad \lambda^{67}_{_{0}}\,\,=\,\,-\frac{1}{2}\left(+\,\lambda_{3} \,-\,\sqrt{3} \lambda_{8}\right),
\end{array}
\end{eqnarray}
and fundementally represents the geometrical relationship between three SU(2) symmetry axes that define a SU(3) group plane.

The three SU(2) groups naturally fall out from the above definitions. Two of the groups, $(\lambda_{1},\lambda_{2},\lambda^{12}_{_{0}})$ and $(\lambda_{6},\lambda_{7},\lambda^{67}_{_{0}})$, are cyclically right-handed, while one group, $(\lambda_{4},\lambda_{5},\lambda^{45}_{_{0}})$, is cyclically left-handed. In right-handed form, the three groups: $(\lambda_{1},\lambda_{2},\lambda^{12}_{_{0}})$, $(\lambda_{5},\lambda_{4},\lambda^{45}_{_{0}})$, and $(\lambda_{6},\lambda_{7},\lambda^{67}_{_{0}})$, each individually follows a cyclic intragroup (same axis) right-handed SU(2) defining algebraic relation $[\lambda_{k},\lambda_{l}]=+2i\lambda_{m}\,\,\epsilon_{klm}$.

In addition to the above intragroup SU(2) algebraic relations, each of the three groups couple to one another through an intergroup (different axis) algebraic relationship: $[\lambda^{a}_{_{0}},\lambda_{l_{1}}]=\pm i\lambda_{l_{2}}$ where the $l_{1}$ and $l_{2}$ subscripts represent two different ``non-0-component'' members from the same (intra-axis) SU(2) group, and $\lambda^{a}_{_{0}}$ is a ``0-component'' member from a different (inter-axis) SU(2) group. A most important relationship between the three groups is $\lambda\,\times\,\lambda\,\,=\,\,0$ for the ``0-component'' matrices:
%\begin{eqnarray}\nonumber\begin{array}{ccc}
\begin{eqnarray}\begin{array}{lll}
\big[\lambda^{12}_{_{0}},\lambda^{45}_{_{0}}\big] = \mathrm{\textbf{0}_{3}}, &
\big[\lambda^{45}_{_{0}},\lambda^{67}_{_{0}}\big] = \mathrm{\textbf{0}_{3}}, &
\big[\lambda^{67}_{_{0}},\lambda^{12}_{_{0}}\big] = \mathrm{\textbf{0}_{3}}.
\end{array}
\end{eqnarray}
The importance of the above commutations means that we have simultaneously measurable eigenvalues along each of the three ``0-component'' quantization axes. As with eigenvector producing two SU(3) eigenvalues, the same eigenvector produces three eigenvalues within the SU(3) symmetry plane, one along each SU(2) symmetry axis contained within the plane.

\section{Multiple Quantization Axes Configurations and Their Symmetries}

The SU(3) group contains three coupled SU(2) symmetry groups that are defined by the (12), (45), and (67) quantization axes. As with any SU(2) symmetry group, each contains a raising operator, a lowering operator, and a ``0-component'' operator. Additionally, since the three SU(2) ``0-component'' matrices commute with one another, simultaneously eigenvalues are not only measurable but basis states are definable along each of the three quantization axes.

This section takes a bottoms up approach in building configurations having a SU(3) group symmetry structure. The only requirement is that all SU(3) configurations are created by basis states defined on one of the three coupled SU(2) symmetry axes within the SU(3) symmetry plane. Even though each SU(2) group is individually algebraically indistinguishable from one another, they are algebraically distinguishable as a geometrically coupled group and therefore fundamentally all multiple quantization axes configurations adhere to the Pauli exclusion principal. Also, the addition of multiple quantization axes produces new symmetry requirements on any eigenvectors of the coupled group.

\subsection{Single Quantization Axis Configurations Within SU(3)}

Using the relation $\tau = \lambda/2$, the above three SU(2) groups each produce intrinsic angular momentum raising, lowering, and ``0-component'' operators. The ``0-component'' operator in turn defines the ``spin-up'' ($\alpha$) and  ``spin-down'' ($\beta$) basis states along the respective quantization axis. In a spherical coordinate system, the three intrinsic angular momentum groups are: ($\tau^{12}_{+},\tau^{12}_{-},\tau^{12}_{_{0}}$), ($\tau^{45}_{+},\tau^{45}_{-},\tau^{45}_{_{0}}$), and ($\tau^{67}_{+},\tau^{67}_{-},\tau^{67}_{_{0}}$) \cite{Pushpa1}.

The first intrinsic spin group within SU(3) is identified with the ``0-component'' operator $\tau^{12}_{_{0}}$ and defined along the (12) quantization axis:
\begin{eqnarray}\nonumber \begin{array}{ccc}
\tau^{12}_{_{+}} & \tau^{12}_{_{-}} & \tau^{12}_{_{0}} \\
\left( \begin{array}{ccc} 0 & -\frac{1}{\sqrt{2}} & 0 \\ 0 & 0 & 0 \\ 0 & 0 & 0\end{array} \right)
& \left( \begin{array}{ccc} 0 & 0 & 0 \\ +\frac{1}{\sqrt{2}} & 0 & 0 \\ 0 & 0 & 0\end{array} \right)
& \left( \begin{array}{ccc} +\frac{1}{2} & 0 & 0 \\ 0 & -\frac{1}{2} & 0 \\ 0 & 0 & 0\end{array} \right),
\end{array}
\end{eqnarray}
where:
\begin{eqnarray}\begin{array}{cc}
\alpha^{12} & \beta^{12} \\
\left( \begin{array}{ccc} 1 \\ 0 \\ 0 \end{array} \right)
& \left( \begin{array}{ccc} 0 \\ 1 \\ 0 \end{array} \right). \\ \\
\end{array}
\end{eqnarray}

The second intrinsic spin group within SU(3) is identified with the ``0-component'' operator $\tau^{45}_{_{0}}$ and defined along the (45) quantization axis:
\begin{eqnarray}\nonumber \begin{array}{ccc}
\tau^{45}_{_{+}} & \tau^{45}_{_{-}} & \tau^{45}_{_{0}} \\
\left( \begin{array}{ccc} 0 & 0 & 0 \\ 0 & 0 & 0 \\ -\frac{1}{\sqrt{2}} & 0 & 0\end{array} \right)
& \left( \begin{array}{ccc} 0 & 0 & +\frac{1}{\sqrt{2}} \\ 0 & 0 & 0 \\ 0 & 0 & 0\end{array} \right)
& \left( \begin{array}{ccc} -\frac{1}{2} & 0 & 0 \\ 0 & 0 & 0 \\ 0 & 0 & +\frac{1}{2}\end{array} \right),
\end{array}
\end{eqnarray}
where:
\begin{eqnarray}\begin{array}{ccc}
\alpha^{45} & \beta^{45} \\
\left( \begin{array}{ccc} 0 \\ 0 \\ 1 \end{array} \right)
& \left( \begin{array}{ccc} 1 \\ 0 \\ 0 \end{array} \right). \\ \\
\end{array}
\end{eqnarray}

The third intrinsic spin group within SU(3) is identified with the ``0-component'' operator $\tau^{67}_{_{0}}$ and defined along the (67) quantization axis:
\begin{eqnarray}\nonumber \begin{array}{ccc}
\tau^{67}_{_{+}} & \tau^{67}_{_{-}} & \tau^{67}_{_{0}}\\
\left( \begin{array}{ccc} 0 & 0 & 0 \\ 0 & 0 & -\frac{1}{\sqrt{2}} \\ 0 & 0 & 0\end{array} \right)
& \left( \begin{array}{ccc} 0 & 0 & 0 \\ 0 & 0 & 0 \\ 0 & +\frac{1}{\sqrt{2}} & 0\end{array} \right)
& \left( \begin{array}{ccc} 0 & 0 & 0 \\ 0 & +\frac{1}{2} & 0 \\ 0 & 0 & -\frac{1}{2}\end{array} \right),
\end{array}
\end{eqnarray}
where:
\begin{eqnarray}\begin{array}{ccc}
\alpha^{67} & \beta^{67} \\
\left( \begin{array}{ccc} 0 \\ 1 \\ 0 \end{array} \right)
& \left( \begin{array}{ccc} 0 \\ 0 \\ 1 \end{array} \right). \\ \\
\end{array}
\end{eqnarray}

The above ``spin-up'' and ``spin-down'' basis eigenvectors shows that $\alpha^{12}$ has the same column matrix structure as $\beta^{45}$, $\beta^{12}$ has the same column matrix structure as $\alpha^{67}$, and $\alpha^{45}$ has the same column matrix structure as $\beta^{67}$.

The algebraic coupling properties of the three intrinsic angular momentum groups within the SU(3) group are as follows:
%\begin{eqnarray}\nonumber\begin{array}{rclcrcl}
\begin{eqnarray}\nonumber \begin{array}{rclc|crcl}
\multicolumn{3}{c}{\mathrm{Intra-Axis}} & & & \multicolumn{3}{c}{\mathrm{Inter-Axis}} \\
\multicolumn{3}{c}{\mathrm{Commutation}\,\,\mathrm{Relations}} & & &
\multicolumn{3}{c}{\mathrm{Commutation}\,\,\mathrm{Relations}}\\ \hline
& & & & & & & \\
\quad \left[\tau^{a}_{_{0}}, \tau^{a}_{_{0}}\right] & = & \,\mathrm{\textbf{0}_{3}},
& \qquad & \qquad  &
\quad \left[\tau^{a}_{_{0}}, \tau^{b}_{_{0}}\right] & = & \,\mathrm{\textbf{0}_{3}},
\\
& & & & & & & \\
\quad \left[\tau^{a}_{_{0}}, \tau^{a}_{_{\pm 1}}\right] & = & \pm\, \tau^{a}_{_{\pm 1}},
& \qquad & \qquad  &
\quad \left[\tau^{a}_{_{0}}, \tau^{b}_{_{\pm 1}}\right] & = & \mp\,\frac{1}{2} \tau^{b}_{_{\pm 1}},
\\
& & & & & & & \\
\quad \left[\tau^{a}_{_{+ 1}},\tau^{a}_{_{- 1}}\right] & = & -\, \tau^{a}_{_{0}},
& \qquad & \qquad  &
\quad \left[\tau^{a}_{_{+ 1}},\tau^{b}_{_{- 1}}\right] & = & \,\mathrm{\textbf{0}_{3}},
\\
& & & & & & & \\
\quad \left[\tau^{a}_{_{\pm1}}, \tau^{a}_{_{\pm1}}\right] & = & \,\mathrm{\textbf{0}_{3}},
& \qquad & \qquad  &
\quad \left[\tau^{a}_{_{\pm1}}, \tau^{b}_{_{\pm1}}\right] & = & -\,\frac{1}{\sqrt{2}} \tau^{c}_{_{\mp1}},
\end{array}
\end{eqnarray}
where ``$a$,'' ``$b$,'' and ``$c$'' are each one of the three cyclic ordered (12:45:67) SU(2) groups contained within SU(3) and $a\neq b\neq c$.\cite{Raphaelian1}

\subsection{Configurations Produced From Two Quantization Axes}

In the coupling of two intrinsic spin 1/2 angular momentum spaces, the coupled spaces form a triplet state configuration with $S_{_{Total}}\,=\,1$ and a singlet state configuration with $S_{_{Total}}\,=\,0$. These states are separated by one unit of angular momentum $\hbar$ along the angular momentum quantization axis. The same methodology applies for the intrinsic spin basis states found in SU(3). Instead of having one group with two basis states, there are three groups with a total of six spin basis states. The two additional SU(2) groups allow for more coupling possibilities than what is available within only one SU(2) group. Additionally, since the ``0-component'' operators commute with one another, we can simultaneously measure basis states along each of the three axes and therefore construct multi-axis configurations with these basis states.

The coupling of two ``spin-up'' basis states allows for the combination of nine algebraically unique basis state pair configurations having an eigenvalue $+1$ within the SU(3) group space. 
\begin{eqnarray}\begin{array}{lll}
\multicolumn{3}{c}{\mathrm{Eigenvalue\,=\,+1}} \\ \\
\alpha^{12}\alpha^{12} \qquad & \qquad 
\alpha^{12}\alpha^{45} \qquad & \qquad
\alpha^{12}\alpha^{67} \\
\alpha^{45}\alpha^{12} \qquad & \qquad
\alpha^{45}\alpha^{45} \qquad & \qquad
\alpha^{45}\alpha^{67} \\
\alpha^{67}\alpha^{12} \qquad & \qquad
\alpha^{67}\alpha^{45} \qquad & \qquad
\alpha^{67}\alpha^{67} 
\end{array}
\end{eqnarray}

Similarly the coupling two ``spin-down'' basis states allows for the combination of nine algebraically unique basis state pair configurations having an eigenvalue $-1$ within the SU(3) group space.
\begin{eqnarray}\begin{array}{lll}
 \multicolumn{3}{c}{\mathrm{Eigenvalue\,=\,-1}} \\ \\
\beta^{12}\beta^{12} \qquad & \qquad
\beta^{12}\beta^{45} \qquad & \qquad
\beta^{12}\beta^{67} \\
\beta^{45}\beta^{12} \qquad & \qquad
\beta^{45}\beta^{45} \qquad & \qquad
\beta^{45}\beta^{67} \\
\beta^{67}\beta^{12} \qquad & \qquad
\beta^{67}\beta^{45} \qquad & \qquad
\beta^{67}\beta^{67}
\end{array}
\end{eqnarray}

Under a symmetry operation in which the spins of the basis state pair are interchanged but the cyclic ordering of coupling axes is held constant, a Local Exchange Transformation [L.E.T.], all of the above configurations are symmetric and therefore are considered spectroscopically as part of a triplet configuration. For example,
\begin{eqnarray}\nonumber\begin{array}{crclc}
\multicolumn{5}{c}{ \mathrm{Local}\,\,\mathrm{Exchange}\,\,\mathrm{Transformation}} \,\, \\
\multicolumn{5}{c}{[\mathrm{L.E.T.}]:\,\,(\uparrow^{a}_{_{1}}\uparrow^{b}_{_{2}}) \,\,\rightleftharpoons\,\,(\uparrow^{a}_{_{2}}\uparrow^{b}_{_{1}}) } \\ \\
& & [\mathrm{L.E.T.}] & & \\
& \Psi\,\,=\,\,[\alpha^{45}\alpha^{12}] & \mapsto & [\alpha^{45}\alpha^{12}] \,\,\, = \,\, \mathbf{+}\,\, [\alpha^{45}\alpha^{12}] \\ \\
& \Psi & \mapsto & \Psi^{'}\,\,=\,\mathbf{+}\,\Psi.
\end{array}
\end{eqnarray}

Under a symmetry operation in which cyclic ordering of coupling axes is interchanged and the spin state follows the axis, a Global Exchange Transformation [G.E.T.], only the six intra-axis (same axis) basis state pairs are symmetric. For example,
\begin{eqnarray}\nonumber\begin{array}{crclc}
\multicolumn{5}{c}{ \mathrm{Global}\,\,\mathrm{Exchange}\,\,\mathrm{Transformation}} \,\, \\ 
\multicolumn{5}{c}{ [\mathrm{G.E.T.}]:\,\,(\downarrow^{a}_{_{1}}\downarrow^{b}_{_{2}})\,\,\rightleftharpoons \,\,(\downarrow^{b}_{_{2}}\downarrow^{a}_{_{1}}) } \\ \\
& & [\mathrm{G.E.T.}] & & \\
& \Psi\,\,=\,\,[\beta^{12}\beta^{12}] & \mapsto & [\beta^{12}\beta^{12}] \,\,\, = \,\, \mathbf{+}\,\, [\beta^{12}\beta^{12}] \\ \\
& \Psi & \mapsto & \Psi^{'}\,\,=\,\mathbf{+}\,\Psi, \\
\multicolumn{5}{l}{ \mathrm{whereas,}}\\ \\
& \Psi\,\,=\,\,[\beta^{12}\beta^{45}] & \mapsto & [\beta^{45}\beta^{12}] \,\,\, \ne \,\, \mathbf{\pm}\,\, [\beta^{12}\beta^{45}] \\ \\
& \Psi & \mapsto & \Psi^{'}\,\,\neq\,\mathbf{\pm}\,\Psi.
\end{array}
\end{eqnarray}

The remaining dozen inter-axis basis state pairs need an additional coupling into a larger configuration before they become eigenstates under this operation. For example, both configurations below are symmetric under a [L.E.T.] symmetry operation while the $\mid 1,-1\big>$ configuration is symmetric and the $\mid 1,+1\big>$ configuration is anti-symmetric under a [G.E.T.] symmetry operation.
\begin{eqnarray}\nonumber\begin{array}{rccc}
\mid 1,-1\big> & = & \multicolumn{2}{l}{ \frac{1}{\sqrt{2}}\,(\beta^{12}\beta^{45}\,+\,\beta^{45}\beta^{12})} \\ \\ 
& & \mathrm{[L.E.T.]} & \quad \mathrm{[G.E.T.]} \\
& \mapsto & +\mid 1,-1\big>, & +\mid 1,-1\big>;  \\ \\ \\
\mid 1,+1\big> & = & \multicolumn{2}{l}{ \frac{1}{\sqrt{2}}\,(\alpha^{67}\alpha^{12}\,-\,\alpha^{12}\alpha^{67})} \\ \\
& & \mathrm{[L.E.T.]} & \quad \mathrm{[G.E.T.]} \\
& \mapsto & +\mid 1,+1\big>, & \quad -\mid 1,+1\big>.
\end{array}
\end{eqnarray}
Interestingly enough the two additional quantization axes within the space defined by the coupled groups enable the creation of configurations that are anti-symmetric under a [G.E.T.]. These anti-symmetric configurations are not available for this spin state configuration when only one quantization axis is available in a group space. The above basis state pairs produce twelve configurations that are both symmetric under [L.E.T.] and [G.E.T.] operations and six configurations that are symmetric under [L.E.T.] but anti-symmetric under [G.E.T.] operations.

There are eighteen distinct coupled basis state pairs that have an eigenvalue within the space equal to zero. None are symmetric or anti-symmetric under local or global exchange transformations:
\begin{eqnarray}\begin{array}{lll}
\multicolumn{3}{c}{\mathrm{Eigenvalue\,=\,0}} \\ \\
\quad \alpha^{12}\beta^{12}, \qquad & \qquad
\quad \alpha^{12}\beta^{45}, \qquad & \qquad
\quad \alpha^{12}\beta^{67}, \\
\quad \alpha^{45}\beta^{12}, \qquad & \qquad
\quad \alpha^{45}\beta^{45}, \qquad & \qquad
\quad \alpha^{45}\beta^{67}, \\
\quad \alpha^{67}\beta^{12}, \qquad & \qquad
\quad \alpha^{67}\beta^{45}, \qquad & \qquad
\quad \alpha^{67}\beta^{67};
\end{array}
\end{eqnarray}
and
\begin{eqnarray}\begin{array}{lll}
\quad \beta^{12}\alpha^{12}, \qquad & \qquad
\quad \beta^{12}\alpha^{45}, \qquad & \qquad
\quad \beta^{12}\alpha^{67}, \\
\quad \beta^{45}\alpha^{12}, \qquad & \qquad
\quad \beta^{45}\alpha^{45}, \qquad & \qquad
\quad \beta^{45}\alpha^{67}, \\
\quad \beta^{67}\alpha^{12}, \qquad & \qquad
\quad \beta^{67}\alpha^{45}, \qquad & \qquad
\quad \beta^{67}\alpha^{67}.
\end{array}
\end{eqnarray}

Intra-axis configurations that are symmetric under both [L.E.T.] and [G.E.T.] have a form similar to $(\alpha^{a}\beta^{a}+\beta^{a}\alpha^{a})$ while configurations that are anti-symmetric under [G.E.T.] have a form similar to $(\alpha^{b}\beta^{b}-\beta^{b}\alpha^{b})$. Inter-axis configurations of the form $(\alpha^{a}\beta^{b}\pm\beta^{a}\alpha^{b})$ are not eigenvectors of [G.E.T.] even though they are eigenvectors of [L.E.T.]. Additionally, inter-axis configurations of the form $(\alpha^{a}\beta^{c}\pm\beta^{c}\alpha^{a})$ are eigenvectors of [G.E.T.] but are not eigenvectors of [L.E.T.]. Eigenvectors that are symmetric under both operations simultaneously have the coupled basis state pairs configured into the following form:
\begin{eqnarray}\nonumber\begin{array}{rcl}
\mid 1,0\big> & = & \frac{1}{2} \left((\alpha^{12}\beta^{45}+\beta^{12}\alpha^{45})+(\alpha^{45}\beta^{12}+\beta^{45}\alpha^{12})\right).
\end{array}
\end{eqnarray}

Similarly, an eigenvector that is symmetric under a [L.E.T.] and anti-symmetric under a [G.E.T] has the form:
\begin{eqnarray}\nonumber\begin{array}{rcl}
\mid 1,0\big> & = & \frac{1}{2}
\left((\alpha^{67}\beta^{12}+\beta^{67}\alpha^{12})-(\alpha^{12}\beta^{67}+\beta^{12}\alpha^{67})\right).
\end{array}
\end{eqnarray}
The eighteen basis pair combinations having a ``0-component'' eigenvalue equal to zero are symmetrized into nine symmetric [L.E.T.] eigenvector configurations, each containing two coupled basis state pairs. The three intra-axis configurations need no additional symmetrization since they are symmetric eigenvectors under both a [L.E.T.] and [G.E.T.] operation. The remaining six inter-axis configurations need an additional symmetrization to produce [G.E.T.] eigenvectors, three of which are [G.E.T.] symmetric and three are [G.E.T.] anti-symmetric. In total, the above basis state pairs produce six $\mid 1,0\big>$ eigenvectors that are both symmetric under [L.E.T.] and [G.E.T.] operations and three configurations that are symmetric under [L.E.T.] but anti-symmetric under [G.E.T.] operations. These eigenvectors are considered as part of the triplet spectroscopic configuration since they are symmetric under a [L.E.T.] operation.

As with the triplet configurations, we can create singlet configurations that are anti-symmetric under an [L.E.T.] operation and have a ``0-component'' eigenvalue within the group space equal to zero. These singlet configurations have the intra-axis forms of $(\alpha^{c}\beta^{c}-\beta^{c}\alpha^{c})$ and inter-axis forms like $(\alpha^{c}\beta^{b}-\beta^{c}\alpha^{b})$. Inter-axis eigenvectors having a definite [G.E.T.] symmetry require additional combinations of these latter configurations. A configuration that is anti-symmetric under [L.E.T.] but symmetric under [G.E.T.] operations has a form similar to:
\begin{eqnarray}\nonumber\begin{array}{rcl}
\mid 0,0\big> & = & \frac{1}{2}
\left((\alpha^{67}\beta^{12}-\beta^{67}\alpha^{12})-(\alpha^{12}\beta^{67}-\beta^{12}\alpha^{67})\right),
\end{array}
\end{eqnarray}
while a configuration that is anti-symmetric under [L.E.T.] and anti-symmetric under [G.E.T.] operations has a form similar to:
\begin{eqnarray}\nonumber\begin{array}{rcl}
\mid 0,0\big> & = & \frac{1}{2}
\left((\alpha^{67}\beta^{12}-\beta^{67}\alpha^{12})+(\alpha^{12}\beta^{67}-\beta^{12}\alpha^{67})\right).
\end{array}
\end{eqnarray}
The eighteen basis pair combinations having a ``0-component'' eigenvalue equal to $0$, are symmetrized into nine anti-symmetric [L.E.T.] eigenvector configurations and therefore are considered as part of the singlet configuration. Three configurations are intra-axis configurations and six are inter-axis configurations.  The three intra-axis group configurations are anti-symmetric eigenvectors under [G.E.T.]. An additional symmetrization of the inter-axis group configurations produces three symmetric eigenvectors and three anti-symmetric eigenvectors under a [G.E.T.] symmetry operation. In total the above basis state pairs produce three $\mid 0,0\big>$ eigenvectors that are anti-symmetric under [L.E.T.] and symmetric under [G.E.T.] operations and six $\mid 0,0\big>$ eigenvectors that are anti-symmetric under both [L.E.T.] and [G.E.T.] operations.

In summary, the two additional SU(2) quantization axes found within SU(3) symmetry group allows for additional coupling possibilities over what is available with only one available SU(2) quantization axis.  Symmetrization of these coupled basis states pairs produce eigenvectors that are symmetric or anti-symmetric under a local exchange transformation. These eigenvectors are similar to those produced from a group with only one SU(2) quantization axis. Due to these extra quantization axes, additional symmetrization produces symmetric and anti-symmetric eigenvectors under a global exchange transformation symmetry operation. Some of these eigenvectors are not available in a space containing only one SU(2) quantization axis.

\subsection{Configurations Involving Three Quantization Axes}

The coupling of three basis states occurs naturally when we ``go with the grain'' defined by the singlet or triplet core configuration. As with  single quantization axis configurations consisting of three spin basis states, the three quantization axes configurations are quartets and doublets. Using ``$a$,'' ``$b$,'' and ``$c$'' as markers for the three cyclic ordered SU(2) groups contained within SU(3), the quartet configurations produced from the triplet core have the form:
\begin{eqnarray}\nonumber\begin{array}{rll}
\mid \frac{3}{2},+\frac{3}{2}\big> & = & (\alpha^{a}\alpha^{a}\alpha^{a}) \quad \mathrm{or} \quad
(\alpha^{a}\alpha^{b}\pm\alpha^{b}\alpha^{a})\alpha^{c}, \\ \\
\mid \frac{3}{2},+\frac{1}{2}\big> & = & \alpha^{a}\alpha^{a}\beta^{a}+(\alpha^{a}\beta^{a}+\beta^{a}\alpha^{a})\alpha^{a} \\
& & \quad \mathrm{or} \quad \alpha^{a}\alpha^{b}\beta^{c}+(\alpha^{a}\beta^{b}+\beta^{a}\alpha^{b})\alpha^{c}, \\ \\
\mid \frac{3}{2},-\frac{1}{2}\big> & = & \beta^{a}\beta^{a}\alpha^{a}+(\alpha^{a}\beta^{a}+\beta^{a}\alpha^{a})\beta^{a} \\
& & \quad \mathrm{or} \quad \beta^{a}\beta^{b}\alpha^{c}+(\alpha^{a}\beta^{b}+\beta^{a}\alpha^{b})\beta^{c}, \\ \\
\mid \frac{3}{2},-\frac{3}{2}\big> & = & (\beta^{a}\beta^{a}\beta^{a}) \quad \mathrm{or} \quad
(\beta^{a}\beta^{b}\pm\beta^{b}\beta^{a})\beta^{c}. \qquad \qquad \qquad
\end{array}
\end{eqnarray}
doublets produced from a triplet core configuration have the form:
\begin{eqnarray}\nonumber\begin{array}{rll}
\mid \frac{1}{2},+\frac{1}{2}\big> & = & +2\alpha^{a}\alpha^{a}\beta^{a}-(\alpha^{a}\beta^{a}+\beta^{a}\alpha^{a})\alpha^{a}, \\
& & \quad \mathrm{or}\,\,\, +2\alpha^{a}\alpha^{b}\beta^{c}-(\alpha^{a}\beta^{b}+\beta^{a}\alpha^{b})\alpha^{c}, \\ \\
\mid \frac{1}{2},-\frac{1}{2}\big> & = & -2\beta^{a}\beta^{a}\alpha^{a}+(\beta^{a}\alpha^{a}\,+\,\alpha^{a}\beta^{a})\beta^{a},\\
& & \quad \mathrm{or}\,\,\, -2\beta^{a}\beta^{b}\alpha^{c}+(\beta^{a}\alpha^{b}+\alpha^{a}\beta^{b})\beta^{c}. \qquad \qquad
\end{array}
\end{eqnarray}
and the doublets produced from a singlet core configuration have the form:
\begin{eqnarray}\nonumber\begin{array}{lll}
\mid \frac{1}{2},+\frac{1}{2}\big> & = & (\alpha^{a}\beta^{a}-\beta^{a}\alpha^{a})\alpha^{a},\\
& & \qquad \mathrm{or}\,\,\, (\alpha^{a}\beta^{b}-\beta^{a}\alpha^{b})\alpha^{c}, \qquad \qquad\\ \\
\mid \frac{1}{2},-\frac{1}{2}\big> & = & (\alpha^{a}\beta^{a}-\beta^{a}\alpha^{a})\beta^{a}, \\
& & \qquad \mathrm{or}\,\,\, (\alpha^{a}\beta^{b}-\beta^{a}\alpha^{b})\beta^{c}. \qquad \qquad
\end{array}
\end{eqnarray}

In total there are nine different definable basis states combinations for each of the above configurations. Three are intra-axis and six are inter-axis. Each of these configurations constitute core configurations and as with the previous two-axis core configurations, extra symmetrization steps are needed to produce [L.E.T.] and [G.E.T.] eigenvectors.

In the strategy to produced fully symmetrized eigenvectors under a [G.E.T.] operation, any first symmetrization step involves the coupling of two cores. This step produces eigenvectors having a limited [G.E.T.] symmetry. Therefore additional symmetrization steps are needed to produce eigenvectors having full a [G.E.T.] symmetry. For example, inter-axis quartet state eigenvectors that are symmetric under a [L.E.T.] and symmetric under a limited [G.E.T.] operation involving the interchange of the ordering of only the first two vector spaces have the form:
\begin{eqnarray}\nonumber\begin{array}{rll}
\mid \frac{3}{2},+\frac{3}{2}\big> & = &
(\alpha^{a}\alpha^{b}\alpha^{c}+\alpha^{b}\alpha^{a}\alpha^{c}), \\ \\
\mid \frac{3}{2},+\frac{1}{2}\big> & = &
(\alpha^{a}\alpha^{b}\beta^{c}+\alpha^{a}\beta^{b}\alpha^{c}+\beta^{a}\alpha^{b}\alpha^{c}) \\
& & \qquad +\,\,(\alpha^{b}\alpha^{a}\beta^{c}+\alpha^{b}\beta^{a}\alpha^{c}+\beta^{b}\alpha^{a}\alpha^{c}), \\ \\
\mid \frac{3}{2},-\frac{1}{2}\big> & = &
(\beta^{a}\beta^{b}\alpha^{c}+\beta^{a}\alpha^{b}\beta^{c}+\alpha^{a}\beta^{b}\beta^{c}) \\
& & \qquad +\,\,(\beta^{b}\beta^{a}\alpha^{c}+\beta^{b}\alpha^{a}\beta^{c}+\alpha^{b}\beta^{a}\beta^{c}), \\ \\
\mid \frac{3}{2},-\frac{3}{2}\big> & = &
(\beta^{a}\beta^{b}\beta^{c}+\beta^{b}\beta^{a}\beta^{c});
\end{array}
\end{eqnarray}
while the inter-axis quartet state eigenvectors that are symmetric under a [L.E.T.] and anti-symmetric under the same limited [G.E.T.] operation have the form:
\begin{eqnarray}\nonumber\begin{array}{rll}
\mid \frac{3}{2},+\frac{3}{2}\big> & = &
(\alpha^{a}\alpha^{b}\alpha^{c}-\alpha^{b}\alpha^{a}\alpha^{c}), \\ \\
\mid \frac{3}{2},+\frac{1}{2}\big> & = &  
(\alpha^{a}\alpha^{b}\beta^{c}+\alpha^{a}\beta^{b}\alpha^{c}+\beta^{a}\alpha^{b}\alpha^{c}) \\
& & \qquad -\,\,(\alpha^{b}\alpha^{a}\beta^{c}+\alpha^{b}\beta^{a}\alpha^{c}+\beta^{b}\alpha^{a}\alpha^{c}), \\ \\
\mid \frac{3}{2},-\frac{1}{2}\big> & = &
(\beta^{a}\beta^{b}\alpha^{c}+\beta^{a}\alpha^{b}\beta^{c}+\alpha^{a}\beta^{b}\beta^{c}) \\
& & \qquad -\,\,(\beta^{b}\beta^{a}\alpha^{c}+\beta^{b}\alpha^{a}\beta^{c}+\alpha^{b}\beta^{a}\beta^{c}), \\ \\
\mid \frac{3}{2},-\frac{3}{2}\big> & = &
(\beta^{a}\beta^{b}\beta^{c}-\beta^{b}\beta^{a}\beta^{c}).
\end{array}
\end{eqnarray}
These latter four anti-symmetric eigenvectors under a limited [G.E.T.] symmetry operation are not available in a space containing only a single quantization axis.

Similarly, the inter-axis doublet configurations emanating from the inter-axis triplet cores produce the following inter-axis configurational forms:
\begin{eqnarray}\nonumber\begin{array}{rll}
\mid \frac{1}{2},+\frac{1}{2}\big> & = &
(+2\alpha^{a}\alpha^{b}\beta^{c}-\alpha^{a}\beta^{b}\alpha^{c}-\beta^{a}\alpha^{b}\alpha^{c}) \\ 
& & \qquad +\,\,(+2\alpha^{b}\alpha^{a}\beta^{c}-\alpha^{b}\beta^{a}\alpha^{c}-\beta^{b}\alpha^{a}\alpha^{c}) \\ \\
\mid \frac{1}{2},-\frac{1}{2}\big> & = &
(-2\beta^{a}\beta^{b}\alpha^{c}+\beta^{a}\alpha^{b}\beta^{c}+\alpha^{a}\beta^{b}\beta^{c}) \\
& & \qquad +\,\,(-2\beta^{b}\beta^{a}\alpha^{c}+\beta^{b}\alpha^{a}\beta^{c}+\alpha^{b}\beta^{a}\beta^{c}),
\end{array}
\end{eqnarray}
and are all symmetric under a [L.E.T.] and a limited [G.E.T.] operation, whereas the doublet configurations
\begin{eqnarray}\nonumber\begin{array}{rll}
\mid \frac{1}{2},+\frac{1}{2}\big> & = &
(+2\alpha^{a}\alpha^{b}\beta^{c}-\alpha^{a}\beta^{b}\alpha^{c}-\beta^{a}\alpha^{b}\alpha^{c}) \\
& & \qquad -\,\,(+2\alpha^{b}\alpha^{a}\beta^{c}-\alpha^{b}\beta^{a}\alpha^{c}-\beta^{b}\alpha^{a}\alpha^{c}) \\ \\
\mid \frac{1}{2},-\frac{1}{2}\big> & = &
(-2\beta^{a}\beta^{b}\alpha^{c}+\beta^{a}\alpha^{b}\beta^{c}+\alpha^{a}\beta^{b}\beta^{c}) \\
& & \qquad -\,\,(-2\beta^{b}\beta^{a}\alpha^{c}+\beta^{b}\alpha^{a}\beta^{c}+\alpha^{b}\beta^{a}\beta^{c}),
\end{array}
\end{eqnarray}
are all symmetric under a [L.E.T.] and anti-symmetric under a limited [G.E.T.] operation. These last two configurations do not exist in a vector space having only one available quantization axis.

Up to this point in the production of symmetrized eigenvectors, the total number possible eigenvectors for each of the four quartet configurations and two doublet configurations equals nine. Six of the nine are symmetric under both a [L.E.T.] and a limited [G.E.T.] operation of which three are intra-axis and three are inter-axis configurations. The remaining three are symmetric [L.E.T.] inter-axis eigenvectors that are anti-symmetric under a limited [G.E.T.] symmetry operation.

Finally, the inter-axis doublet configurations arising from the inter-axis singlet cores produce the following configurational forms that are anti-symmetric under a [L.E.T.] and a limited [G.E.T.] operation:
\begin{eqnarray}\nonumber\begin{array}{rll}
\mid \frac{1}{2},+\frac{1}{2}\big> & = &
(\alpha^{a}\beta^{b}\alpha^{c}-\beta^{a}\alpha^{b}\alpha^{c})\,\,
+\,\,(\alpha^{b}\beta^{a}\alpha^{c}-\beta^{b}\alpha^{a}\alpha^{c}), \\ \\
\mid \frac{1}{2},-\frac{1}{2}\big> & = &
(\beta^{a}\alpha^{b}\beta^{c}-\alpha^{a}\beta^{b}\beta^{c})\,\,
+\,\,(\beta^{b}\alpha^{a}\beta^{c}-\alpha^{b}\beta^{a}\beta^{c}),
\end{array}
\end{eqnarray}
and produce the configurational forms that are anti-symmetric under a [L.E.T.] and symmetric under a limited [G.E.T.] operation:
\begin{eqnarray}\nonumber\begin{array}{rll}
\mid \frac{1}{2},+\frac{1}{2}\big> & = &
(\alpha^{a}\beta^{b}\alpha^{c}-\beta^{a}\alpha^{b}\alpha^{c})\,\,
-\,\,(\alpha^{b}\beta^{a}\alpha^{c}-\beta^{b}\alpha^{a}\alpha^{c}), \\ \\
\mid \frac{1}{2},-\frac{1}{2}\big> & = &
(\beta^{a}\alpha^{b}\beta^{c}-\alpha^{a}\beta^{b}\beta^{c})\,\,
-\,\,(\beta^{b}\alpha^{a}\beta^{c}-\alpha^{b}\beta^{a}\beta^{c}),
\end{array}
\end{eqnarray}
As in the previous case, these last two configurations are not definable within a space containing only one available quantization axis. For a configuration arising from a the singlet core, the total number possible eigenvectors also equals nine for each of the two doublet configurations. Six of the nine are anti-symmetric under both the [L.E.T.] and a limited [G.E.T.] operation. Three are intra-axis and three inter-axis configurations. The remaining three inter-axis configurations are anti-symmetric under a [L.E.T.] operation and symmetric for a limited [G.E.T.] operation.

\subsection{Configurations Involving Four or More Basis States}

Since the SU(3) group contains three SU(2) symmetry quantization axes, the coupling of four or more basis states into configurations now occurs more naturally when we couple the above coupled configurations with one another in a parallel mode using standard vector coupling methods followed by a symmetrization over that of the serial additive mode of coupling additional uncoupled vector spaces to the above configurations. In many ways the four or more basis state parallel coupling methodology mimics the nuclear shell model in which protons and neutrons within the nucleus couple separately prior to any final configuration coupling.

In the parallel coupling mode, coupling singlets or triplets with one another produces configurations containing four basis states. Here, two separate configurations couple together (each of which can be intra-axis or inter-axis configurations). The coupling of core configurations, $[(^{3}S)(^{3}S)]$ produces the terms $^{5}S_{2},\,^{3}S_{1},\,^{1}S_{0}$. The coupling of core configurations involving $[(^{3}S)(^{1}S)]$ produces two $^{3}S_{1}$ states. Finally, the coupling of core configurations $[(^{1}S)(^{1}S)]$ produces the $^{1}S_{0}$ configuration. Additional symmetrization steps are needed to produce the fully symmetrized [L.E.T.] and [G.E.T.] eigenvectors.

In the serial coupling mode, the coupling of a single inter-axis core configuration comprised of three basis states to a single axis (valence-like) basis state produces configurations containing four basis states. Core configurations $([^{3}S]\,^{4}S_{3/2})$, $([^{3}S]\,^{2}S_{1/2})$, and $([^{1}S]\,^{2}S_{1/2})$, coupling to an additional basis state produces produces the terms $^{5}S_{2},\,^{3}S_{1},\,^{1}S_{0}$, $^{3}S_{1}$, and $^{1}S_{0}$. As with the parallel coupling method, extra symmetrization steps produce the fully symmetrized [L.E.T.] and [G.E.T.] eigenvectors.

The transformation from a vector space defined by parallel coupling configuration eigenvectors to one defined by serial coupling configuration eigenvectors occurs through the transformational mapping defined by Clebsch-Gordon coefficient tables. A rotation of tan$\theta=\pm1/\sqrt{2}$ exists between two of the basis states within the coupled space. All other eigenvectors map onto one another singly and therefore represent a $100\%$ geometrical overlap between the two coupling configuration representations. This is shown in the transformation below.
\begin{eqnarray}\nonumber\begin{array}{llll}
\left( \begin{array}{c}
([^{3}S]\,^{4}S_{3/2})\,^{5}S_{2} \\ \\
([^{3}S]\,^{4}S_{3/2})\,^{3}S_{1} \\ \\
([^{3}S]\,^{2}S_{1/2})\,^{3}S_{1} \\ \\
([^{1}S]\,^{2}S_{1/2})\,^{3}S_{1} \\ \\
([^{3}S]\,^{2}S_{1/2})\,^{1}S_{0} \\ \\
([^{1}S]\,^{2}S_{1/2})\,^{1}S_{0}
\end{array} \right)
& = &
M
\left( \begin{array}{c}
([^{3}S](^{3}S))\,^{5}S_{2} \\ \\
([^{3}S](^{1}S))\,^{3}S_{1} \\ \\
([^{3}S](^{3}S))\,^{3}S_{1} \\ \\
([^{1}S](^{3}S))\,^{3}S_{1} \\ \\
([^{3}S](^{3}S))\,^{1}S_{0} \\ \\
([^{1}S](^{1}S))\,^{1}S_{0}
\end{array} \right),
\end{array} \end{eqnarray}
where:
\begin{eqnarray}\nonumber\begin{array}{llll}
M
& = &
\left( \begin{array}{cccccc}
1 & 0 & 0 & 0 & 0 & 0 \\
0 & +\sqrt{\frac{2}{3}} & +\sqrt{\frac{1}{3}} & 0 & 0 & 0 \\ \\
0 & -\sqrt{\frac{1}{3}} & +\sqrt{\frac{2}{3}} & 0 & 0 & 0 \\ \\
0 & 0 & 0 & 1 & 0 & 0 \\ \\
0 & 0 & 0 & 0 & 1 & 0 \\ \\
0 & 0 & 0 & 0 & 0 & 1
\end{array} \right).
\end{array} \end{eqnarray}

As with configurations consisting of four basis states, SU(3) configurations containing five basis states are best created from parallel coupling a doublet or quartet configuration with that of a singlet or triplet configuration. Any combination of inter-axis configurations and intra-axis configurations are possible. The coupling of the core configuration $[(^{4}S_{3/2})(^{3}S_{1})]$ produces the spectral terms of $^{6}S_{5/2}$, $^{4}S_{3/2}$, and $^{2}S_{1/2}$, while the $[(^{4}S_{3/2})(^{1}S_{0})]$ core produces a $^{4}S_{3/2}$ term. Similarly the core configuration of $[(^{2}S_{1/2})(^{3}S_{1})]$ produces both $^{4}S_{3/2}$ and $^{2}S_{1/2}$, while the $[(^{2}S_{1/2})(^{1}S_{0})]$ produces the $^{2}S_{1/2}$ configuration. Finally the core of $[(^{2}S_{1/2})(^{3}S_{1})]$ produces $^{4}S_{3/2}$ and $^{2}S_{1/2}$, and $[(^{2}S_{1/2})(^{1}S_{0})]$ produces $^{2}S_{1/2}$. All of the above configurations need additional symmetrization to become eigenvectors of both a local exchange transformation and a global exchange transformation operations. Clebsch-Gordon coefficient tables details the geometrical overlap between parallel coupling configuration eigenvectors and serial coupling configuration eigenvectors.

\section{Intrinsic Phases Between Multi-Quantization Axes Basis States}

The configurations outlined in the previous section have basis states defined along one of the three intrinsic angular momentum (or isospin) quantization axes found within the SU(3) group. Implicit within these configurations is the concept of the center of the group which itself is governed by the relations found in equation \ref{zero-component}. It is this concept of a group center that leads to two intrinsic properties of the group. The first being a positive directionality associated with each quantization axis as referenced to the group center and defined by the respective $\alpha$ and $\beta$ basis states of the axis. The second being that the quantization axes are geometrically distinguishable within the coupled group. This latter property ensures that for the eigenvectors of the group, no two ``spin-up'' or ``spin-down'' basis states are geometrically equivalent even if their underlying algebraic forms are. In the coupled SU(2) groups of SU(3), the algebraic form of the intrinsic angular momentum $\alpha^{12}$ basis state is structurally equivalent to $\beta^{45}$, the algebraic form of $\beta^{12}$ is structurally equivalent to $\alpha^{67}$, and the algebraic form of $\alpha^{45}$ is the structurally equivalent to $\beta^{67}$.
\begin{eqnarray}\nonumber\begin{array}{lrccc}
& & \mid{} \tau,\tau_{_{0}} \big{>}^{12} & \mid{} \tau,\tau_{_{0}} \big{>}^{45} & \mid{} \tau,\tau_{_{0}} \big{>}^{67} \\ \\
\textrm{``spin-up''} & \alpha \,= & \left( \begin{array}{ccc} 1 \\ 0 \\ 0 \end{array} \right); & \left( \begin{array}{ccc} 0 \\ 0 \\ 1 \end{array} \right); & \left( \begin{array}{ccc} 0 \\ 1 \\ 0 \end{array} \right); \\ \\
\textrm{``spin-down''} & \beta \,= & \left( \begin{array}{ccc} 0 \\ 1 \\ 0 \end{array} \right); & \left( \begin{array}{ccc} 1 \\ 0 \\ 0 \end{array} \right); & \left( \begin{array}{ccc} 0 \\ 0 \\ 1 \end{array} \right).
\end{array}
\end{eqnarray}
The structural equivalence between eigenvectors of one quantization axis with that from another quantization axis permits raising and lowering operators defined along one axis to have the ability to algebraically operate on basis states defined along other axes. This non-zero operational effect on inter-axis basis states coupled with the axes relationship to the center of the group allows for a definite cyclic ordering sequence of the SU(2) quantization axes to be characterized by phase relationships between the axes. This phase is internal to the parameter space defined by the SU(3) group and allows the above structurally equivalent basis states to be distinguishable but operationally physically equivalent to one another. This section explores the internal phase relationships between the groups basis states and the effects of groups operators acting on them.

\subsection{Phase Relationships Between Positive Directionality Quantization Axes Basis States}

As described in the introduction to this section, there is an intrinsic positive directionality associated with the three SU(2) quantization axes that is defined by the respective intrinsic angular momentum $\alpha$ and $\beta$ basis states on the axis. Additionally, the internal cyclic ordering of the three SU(2) quantization axes is defined with respect to the center of the group which itself is governed by the relations found in equation \ref{zero-component}. This cyclic ordering produces a definable geometric phase relation between the three SU(2) quantization axes within the SU(3) group which in turn produces a definable geometric phase between structurally equivalent basis states. The phase equivalence relationships between the basis states contained within the SU(3) group are:
\begin{eqnarray}\begin{array}{lcl}
\alpha^{12}\,\,=\,\,+\,e^{-2\pi/3}\beta^{45}, & \qquad & \beta^{45}\,\,=\,\,+\,e^{+2\pi/3}\alpha^{12}, \\
\alpha^{45}\,\,=\,\,+\,e^{-2\pi/3}\beta^{67}, & \qquad & \beta^{67}\,\,=\,\,+\,e^{+2\pi/3}\alpha^{45}, \\
\alpha^{67}\,\,=\,\,+\,e^{-2\pi/3}\beta^{12}, & \qquad & \beta^{12}\,\,=\,\,+\,e^{+2\pi/3}\alpha^{67}.
\end{array}
\end{eqnarray}

The above phase equivalence between inter-axis basis states allow for another interesting phenomenon to occur. During a raising or lowering operation on a viable inter-axis basis state, the phase associated with the inter-axis basis state is left dangling. This dangling phase is absorbable into any basis states within the configuration (since the phase is internal to the group) to produce configurations having no dangling phases. For example, a lowering operation defined along the (67) axis acting on the ($\beta^{12}\alpha^{45}$) configuration produces the ($\beta^{67}\beta^{67}$) configuration:
\begin{eqnarray}\nonumber\begin{array}{l}
\tau^{67}_{_{-1}}\,(\,\beta^{12} \alpha^{45}\,) \\
\,\,\,\,=\, \tau^{67}_{_{-1}}\,(\,(e^{+2\pi/3}\alpha^{67})\,\alpha^{45}\,) \\
\,\,\,\,=\, \tau^{67}_{_{-1}}\,(\,\alpha^{67}\,(e^{+2\pi/3}\alpha^{45}\,)\,) \\
\,\,\,\,=\, \tau^{67}_{_{-1}}\,(\,\alpha^{67}\,\beta^{67}\,) \,\,=\,\,-\frac{1}{\sqrt{2}}\,(\,\beta^{67}\,\beta^{67}\,),
\end{array}
\end{eqnarray}
where ($\beta^{67}\,\beta^{67}$) is within a phase equivalent to and indistinguishable from the basis state pair configurations of ($\alpha^{45}\,\beta^{67}$), ($\beta^{67}\,\alpha^{45}$), and ($\alpha^{45}\,\alpha^{45}$). The two other lowering operators, $\tau^{12}_{_{-1}}$ and $\tau^{45}_{_{-1}}$, operating on the ($\beta^{12}\alpha^{45}$) configuration produce $\emptyset$ and $-\frac{1}{\sqrt{2}}\,(\beta^{12}\beta^{45})$ respectively. These latter configurations are completely distinguishable from the one that the $\tau^{67}_{_{-1}}$ operator produces. Table \ref{table1} details the resultant intrinsic angular momentum (or isospin) basis states produced from raising and lowering operations on the intra-axis and inter-axis intrinsic angular momentum (or isospin) basis states contained within the SU(3) group.

\begin{table*}
\caption{\label{table1}This table shows the resultant positive directionality defined basis state produced by raising and intra-axis lowering operators acting on positive directionality defined intra-axis and inter-axis intrinsic angular momentum (or isospin) basis states within the SU(3) group. Operations involving inter-axis basis states produce a definable phase (internal to the group) associated with the resultant basis state.}
\begin{tabular}{|c||c|c||c|c||c|c|}
  % after \\: \hline or \cline{col1-col2} \cline{col3-col4} ...
  \hline
  & & & & & & \\
  & $\alpha^{12}$ & $\beta^{12}$ & $\alpha^{45}$ & $\beta^{45}$ & $\alpha^{67}$ & $\beta^{67}$ \\ \hline
  & & & & & & \\
  $\tau^{12}_{_{+1}}$ & $\emptyset$ & $+\frac{1}{\sqrt{2}}\alpha^{12}$ & $\emptyset$ & $\emptyset$ & $+\frac{e^{-2\pi/3}}{\sqrt{2}}\,\alpha^{12}$ & $\emptyset$ \\
  & & & & & &  \\
  $\tau^{12}_{_{-1}}$ & $-\frac{1}{\sqrt{2}}\beta^{12}$ & $\emptyset$ & $\emptyset$ & $-\frac{e^{+2\pi/3}}{\sqrt{2}}\,\beta^{12}$ & $\emptyset$ & $\emptyset$ \\
  & & & & & & \\ \hline \hline
  & & & & & & \\
  $\tau^{45}_{_{+1}}$ & $+\frac{e^{-2\pi/3}}{\sqrt{2}}\,\alpha^{45}$ & $\emptyset$ & $\emptyset$ & $+\frac{1}{\sqrt{2}}\alpha^{45}$ & $\emptyset$ & $\emptyset$ \\
  & & & & & &  \\
  $\tau^{45}_{_{-1}}$ & $\emptyset$ & $\emptyset$ & $-\frac{1}{\sqrt{2}}\beta^{45}$ & $\emptyset$ & $\emptyset$ & $-\frac{e^{+2\pi/3}}{\sqrt{2}}\,\beta^{45}$ \\
  & & & & & & \\ \hline \hline
  & & & & & &  \\
  $\tau^{67}_{_{+1}}$ & $\emptyset$ & $\emptyset$ & $+\frac{e^{-2\pi/3}}{\sqrt{2}}\,\alpha^{67}$ & $\emptyset$ & $\emptyset$ & $+\frac{1}{\sqrt{2}}\alpha^{67}$ \\
  & & & & & &  \\
  $\tau^{67}_{_{-1}}$ & $\emptyset$ & $-\frac{e^{+2\pi/3}}{\sqrt{2}}\,\beta^{67}$ & $\emptyset$ & $\emptyset$ & $-\frac{1}{\sqrt{2}}\beta^{67}$ & $\emptyset$ \\
  & & & & & & \\ \hline
\end{tabular}
\end{table*}

\subsection{Phase Relationships Between Negative Directionality Quantization Axes Basis States}

In some cases these internal phases are associated with basis states definable with respect to the negative directionality of a quantization axis. Within the group, the negative directionality of a quantization axis is orientated $\pm\pi$ relative to the positive directionality of the quantization axis. A ``spin-up'' basis state defined with respect to the positive directionality of a quantization axis is geometrically equivalent to a ``spin-down'' defined with respect to the negative directionality of that quantization axis. Similarly a ``spin-down'' basis state defined with respect to the positive directionality of a quantization axis is equivalent to geometrically a ``spin-up'' defined with respect to the negative directionality of that quantization axis. These negative directionality referenced intrinsic angular momentum basis states have opposite eigenvalues to the basis states defined in reference to the positive directionality of a quantization axis.

The relationship between positive and negative directionality intra-axis basis states is given by the following expressions:
\begin{eqnarray}\begin{array}{lcl}
\overline{\alpha}\,\,=\,\,-\,e^{+\pi}\,\beta & \,\,\mathrm{and}\,\, & \overline{\beta}\,\,=\,\,-\,e^{-\pi}\,\alpha.
\end{array}
\end{eqnarray}

SU(2) eigenvectors found within the SU(3) group referenced to the quantization axes negative directionalities have the following intrinsic angular momentum (or isospin) basis state structural forms:
\begin{eqnarray}\nonumber\begin{array}{lrccc}
& & \mid{} \tau,\tau_{_{0}} \big{>}^{12} & \mid{} \tau,\tau_{_{0}} \big{>}^{45} & \mid{} \tau,\tau_{_{0}} \big{>}^{67} \\ \\
\textrm{``spin-up''} & \overline{\alpha} \,= & \left( \begin{array}{ccc} 0 \\ -1 \\ 0 \end{array} \right); & \left( \begin{array}{ccc} -1 \\ 0 \\ 0 \end{array} \right); & \left( \begin{array}{ccc} 0 \\ 0 \\ -1 \end{array} \right); \\ \\
\textrm{``spin-down''} & \overline{\beta} \,= & \left( \begin{array}{ccc} -1 \\ 0 \\ 0 \end{array} \right); & \left( \begin{array}{ccc} 0 \\ 0 \\ -1 \end{array} \right); & \left( \begin{array}{ccc} 0 \\ -1 \\ 0 \end{array} \right).
\end{array}
\end{eqnarray}
When operated by an intra-axis ``0-component'' operator, the above basis states eigenvectors yield the appropriate eigenvalue for its namesake. A ''spin-up' produces a $+1/2$ eigenvalue while a ''spin-down'' produces a $-1/2$ eigenvalue.

Using the phase equivalences associated with the positive directionality basis states and the relationship between positive and negative directionality intra-axis basis states, the inter-axis relationship between the positive and negative axis orientation group basis states are:
\begin{eqnarray}\begin{array}{lcl}
\alpha^{12}\,\,=\,\,-\,e^{+\pi/3}\,\overline{\alpha^{45}}, & \qquad & \overline{\alpha^{45}}\,\,=\,\,-\,e^{-\pi/3}\,\alpha^{12}, \\
\alpha^{45}\,\,=\,\,-\,e^{+\pi/3}\,\overline{\alpha^{67}}, & \qquad & \overline{\alpha^{67}}\,\,=\,\,-\,e^{-\pi/3}\,\alpha^{45}, \\
\alpha^{67}\,\,=\,\,-\,e^{+\pi/3}\,\overline{\alpha^{12}}, & \qquad & \overline{\alpha^{12}}\,\,=\,\,-\,e^{-\pi/3}\,\alpha^{67},
\end{array}
\end{eqnarray}
and
\begin{eqnarray}\begin{array}{lcl}
\beta^{12}\,\,=\,\,-\,e^{-\pi/3}\,\overline{\beta^{67}}, & \qquad & \overline{\beta^{67}}\,\,=\,\,-\,e^{+\pi/3}\,\beta^{12}, \\
\beta^{45}\,\,=\,\,-\,e^{-\pi/3}\,\overline{\beta^{12}}, & \qquad & \overline{\beta^{12}}\,\,=\,\,-\,e^{+\pi/3}\,\beta^{45}, \\
\beta^{67}\,\,=\,\,-\,e^{-\pi/3}\,\overline{\beta^{45}}, & \qquad & \overline{\beta^{45}}\,\,=\,\,-\,e^{+\pi/3}\,\beta^{67}.
\end{array}
\end{eqnarray}
From the above mapping, the internal phase relationships between the negative directionality basis states are:
\begin{eqnarray}\begin{array}{lcl}
\overline{\alpha^{12}}\,\,=\,\,+\,e^{+2\pi/3}\overline{\beta^{67}}, & \qquad & \overline{\beta^{67}}\,\,=\,\,+\,e^{-2\pi/3}\overline{\alpha^{12}}, \\
\overline{\alpha^{45}}\,\,=\,\,+\,e^{+2\pi/3}\overline{\beta^{12}}, & \qquad & \overline{\beta^{12}}\,\,=\,\,+\,e^{-2\pi/3}\overline{\alpha^{45}}, \\
\overline{\alpha^{67}}\,\,=\,\,+\,e^{+2\pi/3}\overline{\beta^{45}}, & \qquad & \overline{\beta^{45}}\,\,=\,\,+\,e^{-2\pi/3}\overline{\alpha^{67}}.
\end{array}
\end{eqnarray}
These basis states can conceptionally be thought of as being ``anti-basis'' states when viewed from the positive directionality associated with the center of the group.

As with the positive directionality defined basis states, operations involving negative directionality defined basis states are possible. For example, a raising operation defined along the (67) axis acting on the ($\overline{\alpha^{67}}\,\overline{\beta^{45}}$) basis state pair configuration produces the $+\frac{1}{\sqrt{2}}\,(\,\overline{\beta^{67}}\,\overline{\beta^{45}}\,+\,\overline{\beta^{45}}\,\overline{\beta^{67}}\,)$ configuration:
\begin{eqnarray}\nonumber\begin{array}{l}
\tau^{67}_{_{+1}}\,(\,\overline{\alpha^{67}}\,\overline{\beta^{45}}\,) \\
\qquad = \,\,(\,\tau^{67}_{_{+1}}\,\overline{\alpha^{67}}\,) \,\overline{\beta^{45}} \,\,+\,\overline{\alpha^{67}}\,(\,\tau^{67}_{_{+1}}\,\overline{\beta^{45}}\,), \\
\qquad = \,\, (\,+\,\frac{1}{\sqrt{2}}\,\overline{\beta^{67}}\,) \,\overline{\beta^{45}} \,\,
+\,\overline{\alpha^{67}}\,(\,+\frac{1}{\sqrt{2}}e^{-2\pi/3}\overline{\beta^{67}}\,), \\
\qquad = \,\, +\,\frac{1}{\sqrt{2}}\,(\,\overline{\beta^{67}}\,\overline{\beta^{45}}\,) \,\,
+\,\frac{1}{\sqrt{2}}(\,e^{-2\pi/3}\overline{\alpha^{67}}\,)\overline{\beta^{67}}, \\
\qquad = \,\, +\,\frac{1}{\sqrt{2}}\,(\,\overline{\beta^{67}}\,\overline{\beta^{45}}\,) \,\,
+\,\frac{1}{\sqrt{2}}(\,\overline{\beta^{45}}\,)\overline{\beta^{67}}, \\
\qquad = \,\, +\frac{1}{\sqrt{2}}\,(\,\overline{\beta^{67}}\,\overline{\beta^{45}} \,+\,\overline{\beta^{45}}\,\overline{\beta^{67}}\,).
\end{array}
\end{eqnarray}

Table \ref{table2} details the resultant basis states produced from the raising and lowering operations on the negative directionality intra-axis and inter-axis intrinsic angular momentum (or isospin) basis states. A factor of $-1$ can be applied to each of the resultant basis states in table \ref{table2} to transform them into conceptionally equivalent ``anti-basis'' states as referenced to the positive directionality associated with the center of the group. This latter transformation results in an eigenvalue polarity change.

\begin{table*}
\caption{\label{table2} This table shows the resultant negative directionality defined basis state produced by raising and intra-axis lowering operators acting on negative directionality defined intra-axis and inter-axis SU(2) basis states within the SU(3) group. Operations involving inter-axis basis states produce a definable phase (internal to the group) associated with the resultant basis state.}
\begin{tabular}{|c||c|c||c|c||c|c|}
  % after \\: \hline or \cline{col1-col2} \cline{col3-col4} ...
  \hline
  & & & & & & \\
  & $\overline{\alpha^{12}}$ & $\overline{\beta^{12}}$ & $\overline{\alpha^{45}}$ & $\overline{\beta^{45}}$ & $\overline{\alpha^{67}}$ & $\overline{\beta^{67}}$ \\ \hline
  & & & & & & \\
  $\tau^{12}_{_{+1}}$ & $+\frac{1}{\sqrt{2}}\overline{\beta^{12}}$ & $\emptyset$ & $\emptyset$ & $\emptyset$ & $\emptyset$ & $+\frac{e^{-2\pi/3}}{\sqrt{2}}\,\overline{\beta^{12}}$ \\
  & & & & & &  \\
  $\tau^{12}_{_{-1}}$ & $\emptyset$ & $-\frac{1}{\sqrt{2}}\overline{\alpha^{12}}$ & $-\frac{e^{+2\pi/3}}{\sqrt{2}}\,\overline{\alpha^{12}}$ & $\emptyset$ & $\emptyset$ & $\emptyset$ \\
  & & & & & & \\ \hline \hline
  & & & & & & \\
  $\tau^{45}_{_{+1}}$ & $\emptyset$ & $+\frac{e^{-2\pi/3}}{\sqrt{2}}\,\overline{\beta^{45}}$ & $+\frac{1}{\sqrt{2}}\overline{\beta^{45}}$ & $\emptyset$ & $\emptyset$ & $\emptyset$ \\
  & & & & & &  \\
  $\tau^{45}_{_{-1}}$ & $\emptyset$ & $\emptyset$ & $\emptyset$ & $-\frac{1}{\sqrt{2}}\overline{\alpha^{45}}$ & $-\frac{e^{+2\pi/3}}{\sqrt{2}}\,\overline{\alpha^{45}}$ & $\emptyset$ \\
  & & & & & & \\ \hline \hline
  & & & & & &  \\
  $\tau^{67}_{_{+1}}$ & $\emptyset$ & $\emptyset$ & $\emptyset$ & $+\frac{e^{-2\pi/3}}{\sqrt{2}}\,\overline{\beta^{67}}$ & $+\frac{1}{\sqrt{2}}\overline{\beta^{67}}$ & $\emptyset$ \\
  & & & & & &  \\
  $\tau^{67}_{_{-1}}$ & $-\frac{e^{+2\pi/3}}{\sqrt{2}}\,\overline{\alpha^{67}}$ & $\emptyset$ & $\emptyset$ & $\emptyset$ & $\emptyset$ & $-\frac{1}{\sqrt{2}}\overline{\alpha^{67}}$ \\
  & & & & & & \\ \hline
\end{tabular}
\end{table*}

With the above internal phase equivalences in place, operations involving raising and lowering operators can not only transform basis state configurations among different intrinsic angular momentum (or isospin) quantization axes available within the SU(3) group, but also bridge the gap between basis state configurations defined with respect to the positive directionality of the group center and those defined with respect to the negative directionality of the group center.

\begin{eqnarray}\nonumber\begin{array}{l}
\tau^{12}_{_{-}}\,(\,\overline{\alpha^{67}}\,\overline{\alpha^{45}}\,) \\
\qquad = \,\,(\,\tau^{12}_{_{-}}\,\overline{\alpha^{67}}\,) \,\overline{\alpha^{45}} \,\,+\,\overline{\alpha^{67}}\,(\,\tau^{12}_{_{-}}\,\overline{\alpha^{45}}\,), \\
\qquad = \,\,(\,\emptyset\,) \,\overline{\alpha^{45}} \,\,
+\,\overline{\alpha^{67}}\,(\,\tau^{12}_{_{-}}\,(\,e^{+2\pi/3}\,\overline{\beta^{12}}\,)\,), \\
\qquad = \,\, - \frac{1}{\sqrt{2}}\,(\,(\,e^{+\pi/3}\overline{\alpha^{67}}\,)(\,e^{+\pi/3}\overline{\alpha^{12}}\,)\,) \\
\qquad = \,\, -\frac{1}{\sqrt{2}}\,\alpha^{45}\alpha^{67},
\end{array}
\end{eqnarray}
and
\begin{eqnarray}\nonumber\begin{array}{l}
\tau^{45}_{_{+}}\,(\,\overline{\beta^{12}}\,\overline{\beta^{45}}\,) \\
\qquad = \,\, (\,\tau^{45}_{_{+}}\,\overline{\beta^{12}}\,) \,\overline{\beta^{45}} \,\,+\,\overline{\beta^{12}}\,(\,\tau^{12}_{_{+}}\,\overline{\beta^{45}}\,), \\
\qquad = \,\,  (\,+\frac{e^{-2\pi/3}}{\sqrt{2}}\,\overline{\beta^{45}}\,) \,\overline{\beta^{45}}
\,\,+\,\overline{\beta^{12}}\,(\,\emptyset\,), \\
\qquad = \,\,  +\frac{1}{\sqrt{2}}\,(\,(\,e^{-\pi/3}\overline{\beta^{45}}\,)(\,e^{-\pi/3}\overline{\beta^{45}}\,)\,)\\
\qquad = \,\, +\frac{1}{\sqrt{2}}\,\beta^{67}\beta^{67}. \\ \\ \\
\end{array}
\end{eqnarray}

\section{Conclusions}

In conclusion, we have detailed the algebraic structure of the three coupled SU(2) groups contained within the SU(3) group. Each of these three SU(2) groups has an independent quantization axis from which ``spin-up'' and ``spin-down'' intrinsic angular momentum (or isospin) basis states are definable and simultaneously measurable. Since these groups follow standard SU(2) algebraic relations, all the quantum mechanical foundations associated with general angular momentum coupling are valid and applicable to the SU(2) basis states of SU(3). These axes allow for the possibility of both intra-axis basis state configurations (comprised from basis states defined from a single quantization axis) and inter-axis basis state configurations (comprised from basis states defined from multiple quantization axes). Some of these configurations have exchange transformation symmetries that do not exist in vector spaces having only one intrinsic angular momentum (or isospin) quantization axis available for basis state coupling.

Additionally, each of the three quantization axes has an associated positive directionality defined by the two basis states along that axis. Due to the algebraic nature of the coupling of the three equivalent SU(2) algebraic groups into a SU(3) group structure, these previously indistinguishable individual quantization axes (and basis states) become distinguishable by a definable parameter space phase that is geometrically referenceable the center of the coupled group. These phases also enable the definition of a negative directionality to each of the quantization axes. This in turn produces an associated but separate set of basis states to those defined by the positive directionality of the axis. The consequence of these phases is that the intrinsic angular momentum (or isospin) raising and lowering operators not only allow for intra-axis and inter-axis transformations between basis state configurations, but they also allow for transformations between positive and negative quantization axis directionality basis state configurations.

\begin{acknowledgments}
This manuscript has been authored by the National Security Technologies, LLC, under Contract No. DE-AC52-06NA25946 with the U.S. Department of Energy. The United States Government and the publisher, by accepting the article for publication, acknowledge that the United States Government retains a nonexclusive, paid-up, irrevocable, world-wide license to publish or reproduce the published form of this manuscript, or allow others to do so, for United States Government purposes. The U.S. Department of Energy will provide public access to these results of federally sponsored research in accordance with the DOE Public Access Plan (http://energy.gov/downloads/doe-public-access-plan), DOE/NV/25946--2699.
\end{acknowledgments}

\end{document}